\documentclass[12pt]{article}

\usepackage{amsmath,amssymb,exscale}  
\input epsf
\def\gappeq{\mathrel{\rlap {\raise.5ex\hbox{$>$}}
{\lower.5ex\hbox{$\sim$}}}}
\def\permil{$\%\raise.20ex\hbox{$_0$}}
\def\lappeq{\mathrel{\rlap{\raise.5ex\hbox{$<$}}
{\lower.5ex\hbox{$\sim$}}}}

\begin{document}

\baselineskip=20pt

\topmargin -.5cm
\oddsidemargin -0.8cm
\evensidemargin -0.8cm
\pagestyle{empty}
\begin{flushright}
CERN-TH/2000-154
\end{flushright}
\vspace*{5mm}
\begin{center}
{\Large\bf Grand Unified Theories without  the  Desert}
\vspace{2cm}

{\large\bf Alex Pomarol \footnote{On leave of absence from
IFAE, Universitat Aut{\`o}noma de Barcelona, 
E-08193 Bellaterra, Barcelona.}}\\
\vspace{.6cm}
{\it {Theory Division, CERN}\\
{CH-1211 Geneva 23, Switzerland}\\}
\end{center}
\vspace{1cm}
\begin{abstract}
We present a Grand Unified Theory (GUT)
that has GUT fields  with  masses of the order of a TeV,
but at the same time
preserves (at the one-loop level)
the success of  gauge-coupling unification of the MSSM
and  the smallness of proton decay operators.
This scenario 
is based on a  five-dimensional
theory with the extra dimension compactified  as
in the Randall--Sundrum model.
The MSSM gauge sector and its GUT extension live in the 5D bulk,
while the matter sector is localized on a 4D boundary.
\end{abstract}
\vfill
\begin{flushleft}
CERN-TH/2000-154\\
May 2000
\end{flushleft}
\eject
\pagestyle{empty}
\setcounter{page}{1}
\setcounter{footnote}{0}
\pagestyle{plain}


\section{Introduction}

One of the most fascinating challenges in particle physics
is the unification of the forces of nature.
Grand unified theories (GUTs)
have been advocated as an extension of the standard model (SM)
to unify the gauge interactions.
Their immediate consequence is  the explanation of charge quantization.
The scale at which GUTs must replace the SM, however, 
must be very high, $M_{\rm GUT}> 10^{15-16}$ GeV,
in order to avoid  higher-dimensional 
operators that 
would
lead to a large proton decay rate.
Furthermore,
a large $M_{\rm GUT}$ is also needed for gauge-coupling unification.
Since the 4D gauge couplings evolve
logarithmically with the energy scale, 
a large  $M_{\rm GUT}$ is needed 
to allow  the different gauge couplings of the SM to
get closer and, eventually,  unify.
This is the case 
in the supersymmetric extension of the standard model (MSSM)
where gauge couplings actually unify at $M_{\rm GUT}\sim 10^{16}$ GeV.
Since supersymmetry is also needed for the stability of the 
weak scale versus the large GUT scale, 
supersymmetric GUTs  provide a very appealing framework for 
physics beyond the SM.
Nevertheless, we must face the fact that, 
since GUT fields will only appear at very high energies 
$M_{\rm GUT}\sim 10^{16}$ GeV, 
GUTs will never be tested in a direct way.
These theories predict a big ``desert'' from the weak to the GUT scale.

Here  we will present a GUT scenario without  the ``desert''. 
It  has GUT fields  of   masses of the order of a TeV.
This scenario has been  proposed in Refs.~\cite{ap,gp}
and is based on an extension of the MSSM to a five-dimensional
theory with the extra dimension compactified  as
in the Randall--Sundrum model \cite{rs}.
Differently from Ref.~\cite{rs}, however, 
we will consider  the 
MSSM gauge sector and its GUT extension living in 5D
with matter localized on a 4D boundary.
In this letter we will show that, 
even though the theory  is  five-dimensional,
gauge couplings get 
logarithmic corrections at the one-loop level.
Therefore the theory predicts, as in the 4D MSSM
case, 
the right values of the gauge couplings at low energies.
We will also show that 
proton decay operators are suppressed by  
the  high scale $M_{\rm GUT}\sim  10^{16}$ GeV.

Different attempts 
to obtain theories that,
while
predicting  gauge-coupling unification as in the MSSM,
do not have a ``desert'' between the weak and the GUT scale,
can be found in  Refs.~\cite{othera}.

\section{The set-up}

Our set-up is based on the Randall--Sundrum 5D model \cite{rs},
where the bulk is  a slice of 
AdS$_5$.  
This corresponds to 
a 5D non-factorizable geometry with
a the fifth dimension $y$  compactified on an 
orbifold, $S^1/\mathbb{Z}_2$, of radius $R$ with $0\leq y\leq\pi R$. 
The orbifold  fixed points at 
$y^\ast=0$ and $y^\ast=\pi R$ are 4D  boundaries 
of the five-dimensional space-time.
The metric is given by \cite{rs}
\begin{equation}
\label{metric}
        ds^2=e^{-2ky}\eta_{\mu\nu}dx^\mu dx^\nu+dy^2\, ,
\end{equation}
where $1/k$ is the AdS curvature radius and 
$\eta_{\mu\nu}={\rm diag}(-1,1,1,1)$ with $\mu=1,...,4$. 
The fundamental scale in the 5D theory, $M_5$, is related
with the 4D Planck mass, $M_P$, by $M_5^3\simeq kM^2_P$ (for $R>1/k$).
We assume that all the scales are of roughly the same order of
magnitude  $k\simeq M_5\simeq M_P$, with the radius of the extra dimension
slightly larger, $R\sim 11/k$.
The
effective scales on the two boundaries are 
very different. 
On the $y^*=0$ boundary the effective scale is $M_P$,
while
on the $y^*=\pi R$ boundary
this is given by $ke^{-k\pi R}\sim$ TeV (for the assumed value $R\sim 11/k$).
We will hence call these two boundaries
the $M_P$-boundary  and 
the TeV-boundary respectively.

\section{One-loop corrections to the gauge propagator\hfill\\ 
at low energies}

Let us consider a 5D gauge boson  \cite{gauge,ap}, $A_M(x,y)$, 
with $M=(\mu,5)$, 
 living in the  slice of AdS$_5$ described above.
We want to calculate the 
one-loop
corrections  to the  gauge propagator at low energies. 
For simplicity, we will consider a 5D 
scalar QED theory.
We will work in the gauge
$A_5(x,y)=0$, so we only have to consider $A_\mu(x,y)$.
At energies below the Kaluza--Klein (KK) masses, only 
 the massless zero-mode of the photon is relevant. 
This is given by \cite{gauge,ap}
 \begin{equation}
   \label{gauge}
A_\mu(x,y)=\frac{1}{\sqrt{\pi R}} A^{(0)}_\mu(x)+\dots\, .
 \end{equation}
This corresponds to a 4D massless state with a $y$-independent  
wave-function.
Contrary to the graviton case, 
this gauge mode is not localized by the AdS
metric and has the interesting property that it couples to the 
two boundaries of the orbifold with equal strength.

We want to calculate the one-loop corrections to the propagator
of this massless photon
generated by a  5D scalar $\phi$ 
with charge $1$ and
 even under the $\mathbb{Z}_2$.
We will regularize this theory with  a 5D Pauli--Villars (PV) field $\Phi$
of mass $\Lambda$.
This mass  corresponds to the cut-off of the theory
that we will take 
to be $\Lambda\lappeq k$.
Let us decompose the 5D scalar fields $\phi$ and $\Phi$
in  KK modes.
This has  been done in Refs.~\cite{gw,gp}.
For a  5D scalar
particle of mass $M$, the 
approximate KK mass spectrum for $M<k$ 
is given in Table~\ref{table1}.

\begin{table}[h]
\centering
\begin{tabular}{||c|c|c|c||}\hline\hline
&&& \\ 
 & $n=0$ & $n=1$ to $n\lappeq e^{k\pi R}$& 
$n\gappeq e^{k\pi R}$   \\
&&& \\ 
\hline
&&& \\ 
KK masses& $M/\sqrt{2}$ &$\left(n+\frac{\alpha}{2}-\frac{3}{4}\right)\, 
m_{\rm KK}$&
$n\, m_{\rm KK}$  \\
&&& \\ 
\hline\hline
\end{tabular}
\caption{\small Approximate KK mass spectrum 
of a 5D scalar of mass $M$. We have defined
$\alpha=\sqrt{4+\frac{M^2}{k^2}}$ and $m_{\rm KK}=\pi k e^{-k\pi R}$.
We have neglected corrections to the spectrum of ${\cal O}(1/n)$. 
KK modes with masses close to $M$ or $k$
 can have 
deviations of ${\cal O}(1)$ in their masses from the values given here.
Since these deviations only affect 
few KK, they will be neglected.}
\label{table1}
\end{table}

\noindent We have defined the  $n=0$  mode as the mode
that becomes massless in the limit $M\rightarrow 0$.
For $M$ of order $k$, this  mode 
becomes heavier than some of the   KK states since
  $m_{\rm KK}=\pi ke^{-k\pi R}\ll k$ for $R>1/k$.
This is very different from compactifications in a slice of flat space
where the $n=0$  mode (defined as explained above)
is always the lightest state.
This is the effect of the 
AdS$_5$ curvature  that lower the masses of the KK spectrum 
but not the mass of the zero mode.
For the scalar $\phi$ and the PV field $\Phi$, 
we can obtain the KK spectrum using Table~\ref{table1} with
the following values for  $\alpha$:
\begin{equation}
  \label{alphas}
  \alpha^\phi=2\ \  , \ \ \ \ \ 
  \alpha^\Phi=\sqrt{4+\frac{\Lambda^2}{k^2}}\simeq 2+
\frac{\Lambda^2}{4k^2}\, .
\end{equation}
From this KK decomposition we can already infer the magnitude of the quantum
corrections.
For each KK mode of the field $\phi$ there is a KK mode of the PV field
whose mass  acts as a cut-off scale.
Since the masses of the KK  modes
of $\phi$ and $\Phi$ are of the same order
of magnitude, we do not expect large corrections from them.
Nevertheless, the zero mode of the PV field is very heavy, 
${\cal O}(\Lambda)$,
in contrast with the zero mode of $\phi$ that is massless.
Therefore we expect a large correction coming from this large 
mass splitting
of the zero modes 
that will reproduce the quantum corrections of an ordinary 4D theory.

To see this explicitly,
let us now calculate the one-loop contribution from
the scalar $\phi$ to  the propagator of
the photon $A^{(0)}_\mu$.
Defining the  photon self-energy 
by 
$\Pi_{\mu\nu}(q)=[q^2\eta_{\mu\nu}-q_\mu q_\nu]\Pi(q)$,
we find, at zero momentum, that the one-loop contribution is 
given by
\begin{equation}
\label{porpagator}
\Pi(0)
= 
\frac{b_0}{8\pi^2}
\ln\frac{M^\phi_0}{M^\Phi_0}+
\frac{b_{\rm KK}}{8\pi^2}
\sum^{e^{k\pi R}}_{n=1}\ln\frac{M^{\phi}_n}{M^{\Phi}_n}\, ,
\end{equation}
where $M_0^{\phi}$ and 
$M_n^{\phi}$ 
are respectively 
the masses of the zero mode and $n$-KK mode
of the field $\phi$, and similarly for $\Phi$.
We denote by $b_0$ and $b_{\rm KK}$ 
the beta-function coefficients of the zero mode and KK modes respectively.
In the example here we have $b_0=b_{\rm KK}=1/3$.
Using Table~\ref{table1} with Eq.~(\ref{alphas}), we obtain
\begin{equation}
\label{porpagator2}
\Pi(0)\simeq
\frac{b_0}{8\pi^2}
\ln\frac{\mu}{\Lambda}+
\frac{b_{\rm KK}}{8\pi^2}
\int^{e^{k\pi R}}_1 dn\,
\ln\left(\frac{n+1/4}{n+1/4+\Lambda^2/(8k^2)}\right)\, ,
\end{equation}
where as usual we have introduced an infra-red cut-off $\mu$,
and we have
replaced the sum over $n$ by an integral.
Evaluating the integral and considering that $\Lambda\lappeq k$,
we obtain
\begin{equation}
\Pi(0)
\simeq 
\frac{b_0}{8\pi^2}
\ln\frac{\mu}{\Lambda}+
\frac{b_{\rm KK}}{64\pi^2}\frac{\Lambda^2}{k^2}
\ln\frac{5 m_{\rm KK}}{4\pi k}\, .
\end{equation}
For $\mu\ll\Lambda< k$
the KK contribution 
is small and can be neglected.
We then obtain that the contribution to the gauge boson propagator
is dominated by the zero mode and gives exactly the same 
contribution as in 4 dimensions:
\begin{equation}
\Pi(0)
\simeq 
\frac{b_0}{8\pi^2}
\ln\frac{\mu}{\Lambda}\, .
\label{fr}
\end{equation}
This is the main result of this paper.
It shows that in these 5D theories the contribution to 
the massless-mode
gauge-boson 
propagator 
depends logarithmically on the high-energy cut-off $\Lambda$,
provided that  $\mu\ll\Lambda\lappeq k$.

It is important to see  if the result
of Eq.~(\ref{fr}) can be understood as a
running of the
gauge coupling similar to the  4D case. 
Of course, this cannot be the case if matter is localized on
the TeV-boundary, since on that boundary our effective scale
(cut-off scale) is TeV, and above this energy
effects must be considered from  the fundamental (string) theory. 
Nevertheless, if matter is localized on the $M_P$-boundary,
where the effective scale is $M_P$,
we will show in the next section that 
the effective gauge couplings can be considered
to run logarithmically with the energy similarly 
to the  4D case.

\section{The 5D gauge propagator at high energies}

In order to understand what is the behavior of the
theory at energies above the TeV, we will 
derive here the 5D propagator of the gauge boson 
 in  the AdS$_5$ slice.
Since we are interested in the propagator at high energies 
from the point of view of the $M_P$-boundary, 
 we will consider  the limit $R\rightarrow\infty$.
In this limit we do not need to do KK decomposition
and we can work directly in 5D.

Let us take  the gauge $A_5(x,y)=0$
and consider only the transverse part of $A_\mu(x,y)$, {\it i.e.}
we impose $\partial^\mu A_\mu(x,y)=0$.
It is shown in Ref.~\cite{liu} that the transverse  part 
decouples from the non-transverse
part in the equations of motion.
Moreover, only the propagator of the transverse  part is relevant
to  sources localized on the $M_P$-boundary 
since the current there is transverse, $\partial^\mu J_\mu=0$.
Following   similar steps 
to those in the graviton case \cite{gt,gkr},
we want to calculate the Green function 
for the gauge boson  defined as
\begin{equation}
  \label{def}
A_\mu(x,y)=\int d^4x^\prime dy^\prime \sqrt{-g}\,  
G(x,y,;x^\prime,y^\prime)  
e^{2ky^\prime} J_\mu(x^\prime, y^\prime)\, ,
\end{equation}
with  $\partial^\mu J_\mu=0$.
It can be shown that 
 $e^{k(y+y^\prime)} G(x,y;x^\prime,y^\prime)$
is also 
the Green function of a scalar with mass $-3k^2+2k\delta(y)$ \cite{gp}.
Let us 
change the extra dimensional coordinate to
 $z=e^{ky}/k$.
Taking the 4D Fourier transform of the Green function
\begin{equation}
  \label{Fourier}
 G(x,z;x^\prime,z^\prime)=\int \frac{d^4p}{(2\pi)^4}e^{ip(x-x^\prime)}
G_p(z,z^\prime)\, ,
\end{equation}
we have that $G_p(z,z^\prime)$  must satisfy the equation
\begin{equation}
  \label{propaeq}
\Bigg[\partial^2_z-\frac{1}{z}\partial_z-p^2 \Bigg] G_p(z,z^\prime)=zk
\delta(z-z^\prime)\, .
\end{equation}
Solving Eq.~(\ref{propaeq}) with the Neumann boundary condition
on the $M_P$-boundary,  we find
\begin{equation}
  \label{propa}
 G(x,z;x^\prime,z^\prime)=\frac{i\pi k}{2}zz^\prime
\int \frac{d^4p}{(2\pi)^4}e^{ip(x-x^\prime)}
\Bigg[\frac{J_0(ip/k)}{H^{(1)}_0(ip/k)}H^{(1)}_1(ipz)H^{(1)}_1(ipz^\prime)-
J_1(ipz_<)H^{(1)}_1(ipz_>)\Bigg]\, ,
\end{equation}
where
$H_\nu^{(1)}=J_\nu+iY_\nu$  is the Hankel function of order $\nu$,
 $J_\nu$ and $Y_\nu$  are  Bessel functions, 
and we have defined
$z_>$ ($z_<$) as the greater (lesser) of $z$ and $z^\prime$.
In the case where the coordinate $z^\prime$  is on the $M_P$-boundary,
$z^\prime=1/k$, the Green function simplifies to
\begin{equation}
  \label{propaM_P}
 G(x,z;x^\prime,\frac{1}{k})=-ikz
\int \frac{d^4p}{(2\pi)^4}e^{ip(x-x^\prime)}\frac{1}{p}
\frac{H^{(1)}_1(ipz)}{H^{(1)}_0(ip/k)}\, .
\end{equation}
We are interested in the limit
 $r=|x-x^\prime|\gg 1/k$ where 
 the Green function
is dominated by the  small-momentum values of the Fourier transform 
($p\ll k$) and 
the Hankel function $H^{(1)}_0(ip/k)$ is approximately
$H^{(1)}_0(ip/k)=i\frac{2}{\pi}[\ln\frac{p}{2k}+\gamma]
+{\cal O}(\frac{p^2}{k^2})$.
 
Let us now study  the Green function of Eq.~(\ref{propaM_P}) 
in two different limits. 
First 
let us consider  the propagator at large $z$, $z\gg r$.
In this case we find a falloff of the Green function
\begin{equation}
  \label{propalimitA}
 G(x,z\gg r;x^\prime,\frac{1}{k})\sim\frac{k}{z^2\ln(kz)}\, ,
\end{equation}
that is similar to the graviton case \cite{gt,gkr}.
Eq.~(\ref{propalimitA}) implies  that at large momentum
$p\gappeq 1/z=ke^{-ky}$ 
the $M_P$-boundary decouples from the TeV-boundary. 
This is, in fact,  a check of   consistency of the theory
that shows that the two different
effective scales 
on the boundaries can coexist.

Let us now consider the opposite limit, $r\gg z$, that 
also corresponds to the case when $z=1/k$, {\it i.e.}
the gauge propagator on the $M_P$-boundary. 
In this case we have,
\begin{equation}
  \label{propalimitB}
 G(x,\frac{1}{k};x^\prime,\frac{1}{k})\simeq k
\int \frac{d^4p}{(2\pi)^4}e^{ip(x-x^\prime)}\frac{1}{p^2\ln(p/k)}\, .
\end{equation}
From Eq.~(\ref{propalimitB}) we can derive
the static potential on the $M_P$-boundary:
\begin{equation}
V(r)\sim \frac{k}{r}\, \frac{1}{\ln(kr)}\, .
\end{equation}
We see that it differs from the  Coulomb potential in 4D 
by a 
logarithmic
term.
It means that the gauge coupling on the $M_P$-boundary
grows,  at the tree-level,
logarithmically with the energy.
This is in contrast with  5D theories in 
flat space  where at the classical level the 
coupling grows linearly with the energy.
In a theory with finite $R$ this ``running'' will be present at energies
above the mass of the first KK mode  
$\sim m_{\rm KK}=\pi ke^{-k\pi R}$ (below $m_{\rm KK}$
we have a single massless gauge  boson as in 4D). 
We then have
\begin{equation}
  \label{classru}
g^2(p>m_{\rm KK})\simeq g^2(p\sim m_{\rm KK})\frac{ \pi kR}{\ln{k/p}}\, .
\end{equation}
This mild logarithmic evolution of the gauge coupling 
allows us to go to high energies without entering in the strong
coupling regime.
We must stress that 
this   ``running'' of the gauge coupling
is a tree-level effect,
not a quantum one.
As a consequence, it will be universal for the different
groups of the SM and will not affect gauge-coupling unification.

From the tree-level behaviour of the propagator in Eq.~(\ref{propa}),
we learn
that the theory remains weakly coupled 
for $p\lappeq 1/z$. This suggests that
the theory can be renormalized as long as
we keep our cut-off scale below $1/z$, {\it i.e.}
$\Lambda\lappeq ke^{-ky}$.
Notice that this cut-off scale depends on the position  in the
extra dimension. This should be expected 
in a  theory 
with the metric (\ref{metric}), since
the effective scale of the  4D space-time 
at the position $y$ 
is given by  $\sim ke^{-k y}$.
Using the cut-off $\Lambda=ke^{-ky}$, we can calculate
quantum corrections in a very simple way. 
We just need the 5D propagators for $r>z$.
For the case of the 5D massless scalar discussed in the previous section,
the propagator  behaves as in 4D
\cite{gt,gkr}, $k\int d^4p/(2\pi)^4 (1/p^2)$,
giving then the same one-loop correction to the
gauge coupling as in 4D.

\section{GUTs in a slice of AdS$_5$}

Let us now proceed 
to show that
theories with gauge bosons
in a slice of AdS$_5$ 
can have gauge-coupling unification. 
We will take a top--down approach.
We will assume that we have a supersymmetric
GUT in the slice of AdS$_5$
and 
show that  this theory, when broken
to the MSSM group,
leads to a successful prediction for the 
gauge couplings at low energies.

As a toy example, let us consider an SU(5) theory.
Due to the $\mathbb{Z}_2$ orbifold symmetry, 
the massless gauge sector of this theory consists of a
$N=1$ vectormultiplet \cite{gp}.
They contain the SM gauge bosons plus the GUT gauge bosons, $X$ and 
$Y$, that complete the SU(5) representation.
The KK spectrum consists of $N=2$ vectormultiplets
with masses  $\sim (n-\frac{1}{4})\, m_{\rm KK}$.
Let us now consider that on the $M_P$-boundary  we have 
a chiral supermultiplet, in the {\bf 24} representation of SU(5), 
whose scalar gets a vacuum expectation value (VEV) 
equal to $M_{\rm GUT}\sim 10^{16}$ GeV
(slightly below $k\simeq M_P$) breaking the SU(5) group down 
to the MSSM. 
This can be achieved in the same way as in ordinary 4D SU(5) theories,
since our theory on the boundary is 4D
$N=1$ supersymmetric.
It is easy to calculate the KK spectrum of the resulting theory. 
The $n=0$ MSSM gauge bosons remain massless, while the 
GUT gauge bosons, $X$ and $Y$, have masses
$M_{X,Y}\simeq M_{\rm GUT}$ \footnote{
More precisely, we find  that
$M_{X,Y}$
is determined  by the equation
 $M_{X,Y}(\ln(M_{X,Y}/2k)+\gamma+1/2)+M_{\rm GUT}^2/M_{X,Y}=0$.}. 
The KK mass spectrum  ($n\geq 1$), however, is not modified 
by the VEV of the ${\bf 24}$
(up to corrections of ${\cal O}(M^2_n/k^2)$)
and therefore the KK modes 
approximately respect the  SU(5) symmetry. 
Consequently, 
only the zero modes (as we claimed before)
will give a relative one-loop contribution 
to the SM gauge couplings which, at energies $\mu$,
is given by
\begin{equation}
\frac{1}{\alpha_{i}}=
\frac{1}{\alpha_j}+\frac{b_{i}-b_{j}}{2\pi}
\ln\frac{M_{\rm GUT}}{\mu}\, ,
\end{equation}
where $b_{i}$ 
is the contribution of the massless modes
to the beta-function coefficient of the MSSM gauge group $i$.
Therefore, in order to have the same
predictions for the gauge coupling as in 4D supersymmetric GUTs,
we must just demand that  the massless states of the theory 
be those of the 
MSSM.
This will be the case of the gauge sector, as we already explained.
For the  Higgs sector we can, as usual, assume that they arise from
a  ${\bf 5}$ and $\bar{\bf 5}$ of SU(5).
Since we need to have only the SU(2)$_L$-doublet light,
we will need a mechanism that provides a 
doublet--triplet mass splitting inside the ${\bf 5}$ and $\bar{\bf 5}$.
Several mechanisms exist in the literature for 4D.
It is not clear if these mechanisms can also work in 5D.
Nevertheless, we can  just rely on these
mechanisms by assuming that the Higgs live on the $M_P$-boundary.

Finally, we must implement the matter sector. 
Since they form complete SU(5) multiplets, $\bar{\bf 5}$ and ${\bf 10}$,
they are irrelevant to gauge-coupling unification
(they will  not contribute at the one-loop level to the relative corrections
to the gauge couplings).
The matter sector, however, must satisfy
important constraints 
from proton decay.
Since there are very light
 KK modes of the $X$ and $Y$ bosons ($m_{\rm KK}\sim$
TeV),
we must worry about 
proton-decay operators induced by these modes.
If we analyze the $y$-dependent wave-function of these modes, 
however,
we find that they are peaked on the TeV-boundary \cite{ap}. 
Therefore, 
proton-decay constraints can  be satisfied 
by just placing the matter sector on the $M_P$-boundary.
In this case,
even if we sum over the full KK tower of the 
$X$ and $Y$
bosons, we obtain that the strength of the 
 dimension-six proton-decay operator is given by
\begin{equation}
  \label{eq:pd}
 \sum_n g^2_n\frac{1}{m^2_n}\simeq g^2\frac{\pi k R}{M^4_{\rm GUT}}
\int^{M_{\rm GUT}}_0 m\, dm=g^2\frac{\pi k R}{2M^2_{\rm GUT}}\, ,
\end{equation}
where $g_n$ and $m_n$ are respectively the coupling to the $M_P$-boundary 
and the mass of the KK of the $X,Y$
bosons  that can be derived following
 Ref.~\cite{gp}. 
We see that the result is similar to that in a 4D theory 
where one finds $g^2/M^2_{\rm GUT}$.
The operator (\ref{eq:pd}) is, however, slightly larger in our theory
than in 4D theories
because of the   factor $\pi k R$ in Eq.~(\ref{eq:pd}).
This enhancement is due to the fact that the gauge coupling
grows (at tree-level) 
with the energy according to Eq.~(\ref{classru}).
Notice that,
at the scale $M_{\rm GUT}$,
the theory is close to the strong coupling regime.
This is why 
we  expect in these GUTs 
a proton decay rate for $p\rightarrow \pi e$
closer to the experimental limit than in 4D GUTs.

Up to now we have just assumed that $m_{\rm KK}$ 
(approximately the mass of the lightest KK state) is an independent
parameter
of the theory that we have taken to be close to the 
weak scale by choosing $R\simeq 11/k$.
Nevertheless, it would be  interesting to relate  $m_{\rm KK}$
with the weak scale.
One way to do this is by 
associating the supersymmetry-breaking scale with $m_{\rm KK}$.
A realization of this is given in Ref.~\cite{gp}.
By assuming different boundary conditions 
for bosons and fermions on the TeV-boundary, 
we can get a fermion--boson mass splitting
of  ${\cal O} (m_{\rm KK})$.
This breaks supersymmetry and induces
a Higgs mass of
${\cal O} (m_{\rm KK})$ \cite{gpt}. If this mass is negative,
this will trigger electroweak symmetry breaking.
This scenario therefore links the scale 
$m_{\rm KK}$ with the weak scale.

\subsection{GUT physics at  TeV energies}

Although this theory resembles the ordinary 4D supersymmetric GUT,
it has very different  implications at TeV-energies.
While  4D supersymmetric GUTs predict that only the MSSM fields
have masses of the order of the weak scale, with a big ``desert''
up to the GUT scale,
our theory has plenty of new physics at the TeV.
There are the KK states not only of the SM but also of the GUT fields
and graviton.
It has been shown in Ref.~\cite{ap} that
the KK modes 
of the SM gauge bosons have sizeable couplings to the SM fermions
living on the $M_P$-boundary ($\simeq 0.2 g$) and therefore
they could be seen as  resonances in TeV colliders.
On the other hand, the KK modes of the GUT fields couple very weakly to 
the $M_P$-boundary (this is why the proton decay rate is small).
These modes, however, can be produced at TeV energies
by processes mediated by
 virtual SM gauge bosons 
(that live in the 5D bulk and propagate
between the two boundaries).
At these energies, also   graviton KK modes can be produced.  
In fact, since the effective scale on 
the TeV-boundary
is
$\sim ke^{-k\pi R}\sim$ TeV, quantum gravity or string effects 
can be important and possible to test.

\section{Conclusion}

We have shown that, in theories with 
gauge  bosons propagating  in the 5D bulk of the Randall--Sundrum model
(a slice of  AdS$_5$), the gauge coupling   
gets logarithmic corrections similar to those  in 4D.
These theories contain light (TeV) KK excitations,
but only the (massless) zero modes
contributes to the renormalization of the gauge coupling.

We have proposed 
a GUT
where the gauge bosons live in  the 5D bulk, while
matter is   localized on the $M_P$-boundary.
The theory has the  massless spectrum of the MSSM and 
predicts  
the right value for the gauge couplings at low energies.
On the other hand,
we find, at the TeV scale,
KK modes of the GUT fields.
These modes are very weakly coupled to the fermions of the SM
and consequently
proton decay rates  are  suppressed.
Nevertheless, they couple to the SM gauge bosons with sizeable
couplings, providing the possibility to test 
GUTs at TeV colliders.

\section*{Acknowledgements}

We thank 
Jaume Garriga, 
Tony Gherghetta and 
Riccardo Rattazzi
for very useful discussions.
This work has been supported in part by the Spanish CICYT 
contract AEN99-0766.

\end{document}